%
%
%
\font\ninerm=cmr9 
\font\ninei=cmmi9
\font\nineit=cmti9
\font\ninesl=cmsl9
\font\ninebf=cmbx9
\font\ninesy=cmsy9
\def\rmnine{\fam0\ninerm}
\def\itnine{\fam\itfam\nineit}
\def\slnine{\fam\slfam\ninesl}
\def\bfnine{\fam\bffam\ninebf}
\def\ninepoint{\let\rm=\rmnine
\textfont0=\ninerm \scriptfont0=\sevenrm \scriptscriptfont0=\fiverm
\textfont1=\ninei\scriptfont1=\seveni \scriptscriptfont1=\fivei
\textfont2=\ninesy
\textfont\itfam=\nineit \let\it=\itnine
\textfont\slfam=\ninesl \let\sl=\slnine
\textfont\bffam=\ninebf \scriptfont\bffam=\sevenbf
\scriptscriptfont\bffam=\fivebf
\let\bf=\bfnine
\let\sc=\sevenrm
\normalbaselineskip=11pt\normalbaselines\rm}
\font\tenib=cmmib10
\font\tensc=cmcsc10
\def\rmten{\fam0\tenrm}
\def\itten{\fam\itfam\tenit}
\def\slten{\fam\slfam\tensl}
\def\bften{\fam\bffam\tenbf}
\def\tenpoint{\let\rm=\rmten
\textfont0=\tenrm\scriptfont0=\sevenrm\scriptscriptfont0=\fiverm
\textfont1=\teni\scriptfont1=\seveni\scriptscriptfont1=\fivei
\textfont2=\tensy
\textfont\itfam=\tenit \let\it=\itten
\textfont\slfam=\tensl \let\sl=\slten
\textfont\bffam=\tenbf
\scriptfont\bffam=\sevenbf
\scriptscriptfont\bffam=\fivebf
\let\bf=\bften
\let\sc=\tensc
\normalbaselineskip=12pt\normalbaselines\rm}
\font\twelmib=cmmib10 scaled \magstep1
\font\twelbf=cmbx10   scaled \magstep1
\font\twelsy=cmsy10   scaled \magstep1
\def\rmbftwel{\fam0\twelbf}
\def\itbftwel{\fam\itfam\twelmib}
\def\bftwel{\fam\bffam\twelbf}
\def\bftwelpoint{\let\rm=\rmbftwel
\textfont0=\twelbf  \scriptfont0=\tenbf \scriptscriptfont0=\ninebf
\textfont1=\twelmib \scriptfont1=\tenib \scriptscriptfont1=\tenib
\textfont2=\twelsy
\textfont\itfam=\twelmib  \let\it=\itbftwel
\textfont\slfam=\twelmib  \let\sl=\itbftwel 
\textfont\bffam=\twelbf   \scriptfont\bffam=\tenbf
\scriptscriptfont\bffam=\ninebf
\let\bf=\bftwel
\normalbaselineskip=14pt\normalbaselines\rm}
\font\fortmib=cmmib10 scaled \magstep2
\font\fortbf=cmbx10   scaled \magstep2
\font\fortsy=cmsy10   scaled \magstep2
\def\rmbffort{\fam0\fortbf}
\def\itbffort{\fam\itfam\fortmib}
\def\bffort{\fam\bffam\fortbf}
\def\bffortpoint{\let\rm=\rmbffort
\textfont0=\fortbf  \scriptfont0=\twelbf \scriptscriptfont0=\tenbf
\textfont1=\fortmib \scriptfont1=\twelmib \scriptscriptfont1=\tenib
\textfont2=\fortsy
\textfont\itfam=\fortmib  \let\it=\itbffort
\textfont\slfam=\fortmib  \let\sl=\itbffort 
\textfont\bffam=\fortbf   \scriptfont\bffam=\twelbf
\scriptscriptfont\bffam=\tenbf
\let\bf=\bffort
\normalbaselineskip=16pt\normalbaselines\rm}
%
\def\ul#1{$\setbox0=\hbox{#1}\dp0=0pt\mathsurround=0pt
\underline{\box0}$}
%
\pageno=1
\parindent=4mm
\hsize=120mm \hoffset=20mm
\vsize=190mm \voffset=20mm
\outer\def\bye{\bigskip\vfill\supereject\end}
%
\def\raggedcenter{\leftskip=4em plus 12em \rightskip=\leftskip
  \parindent=0pt \parfillskip=0pt \spaceskip=.3333em \xspaceskip=.5em
  \pretolerance=9999 \tolerance=9999
  \hyphenpenalty=9999 \exhyphenpenalty=9999 }
\def\\{\break}
\newtoks\LECTURE \LECTURE={LECTURE}
\newcount\CNlecture \CNlecture=1
\long\def\title#1{\bgroup\vglue10mm\raggedcenter
	             {\tenpoint\bf\the\LECTURE\ \ArabicCN{lecture}}
	             \vskip10mm
	             {\bffortpoint #1}}
\long\def\author#1{\vskip5mm\tenpoint\rm #1}
\def\inst#1{$(^{#1})$}
\long\def\institute#1{\vskip5mm\tenpoint\sl#1}
\def\maketitle{\vskip5mm
               \vtop{\baselineskip=4pt
                     \vrule height.5pt width3cm\par
                     \vrule height.5pt width2cm}
               \vskip20mm\egroup\tenpoint
	          \let\lasttitle=Y\everypar={\let\lasttitle=N}}
%
\newbox\boxtitle
\newskip\beforesect \newskip\aftersect 
\newskip\beforesubsect \newskip\aftersubsect 
\newskip\beforesubsubsect \newskip\aftersubsubsect
%
\beforesect=7mm plus1mm minus1mm   \aftersect=5mm plus.5mm minus.5mm
\beforesubsect=5mm plus.5mm minus.5mm \aftersubsect=3mm plus.2mm minus.1mm 
\beforesubsubsect=3mm plus.2mm minus.1mm \aftersubsubsect=2mm plus.2mm minus.1mm 
\newcount\CNTa \CNTa=0
\newcount\CNTb \CNTb=0
\newcount\CNTc \CNTc=0
\newcount\CNTd \CNTd=0
\def\resetCN #1{\global\csname CN#1\endcsname =0}
\def\stepCN #1{\global
\expandafter\advance \csname CN#1\endcsname by 1}
\def\ArabicCN #1{\expandafter\number\csname CN#1\endcsname}
\def\RomanCN #1{\uppercase\expandafter{\romannumeral\csname CN#1\endcsname}}
\newcount\sectionpenalty  \sectionpenalty=0
\newcount\subsectionpenalty  \subsectionpenalty=0
\newcount\subsubsectionpenalty  \subsubsectionpenalty=0
\newdimen\indsect
\newdimen\dimensect
\indsect=1cm
\dimensect=\hsize\advance\dimensect by -\indsect
\def\section#1#2 {\par\resetCN{Tb}\resetCN{Tc}\resetCN{Td}%
              \if N\lasttitle\else\vskip-\beforesect\fi
              \bgroup
              \bf
              \pretolerance=20000
              \setbox0=\vbox{\vskip\beforesect
                       \noindent\ArabicCN{Ta}.\kern1em#1#2
                       \vskip\aftersect}
              \dimen0=\ht0\advance\dimen0 by\dp0 
              \advance\dimen0 by 2\baselineskip
              \advance\dimen0 by\pagetotal
              \ifdim\dimen0>\pagegoal
                 \ifdim\pagetotal>\pagegoal
                 \else\eject\fi\fi
              \vskip\beforesect
              \penalty\sectionpenalty \global\sectionpenalty=-200
              \global\subsectionpenalty=10007
              \ifx#1*\noindent #2\else\stepCN{Ta}
              \setbox0=\hbox{\noindent\ArabicCN{Ta}.}
              \indsect=\wd0\advance\indsect by 1em
              \parshape=2 0pt\hsize \indsect\dimensect
              \noindent\hbox to \indsect{\ArabicCN{Ta}.\hfil}#1\fi
              \vskip\aftersect
              \egroup
              \let\lasttitle=Y
              \nobreak\parindent=0pt
              \everypar={\parindent=4mm
                         \penalty0\let\lasttitle=N}\ignorespaces}
\def\subsection#1#2 {\par\resetCN{Tc}\resetCN{Td}%
              \if N\lasttitle\else\vskip-\beforesubsect\fi
              \bgroup\tenpoint\bf
                 \setbox0=\vbox{\vskip\beforesubsect
                 \noindent\ArabicCN{Ta}.\ArabicCN{Tb}.\kern1em#1#2
                 \vskip\aftersubsect}
              \dimen0=\ht0\advance\dimen0 by\dp0\advance\dimen0 by
                 2\baselineskip
              \advance\dimen0 by\pagetotal
              \ifdim\dimen0>\pagegoal
                 \ifdim\pagetotal>\pagegoal
                 \else \if N\lasttitle\eject\fi \fi\fi
              \vskip\beforesubsect
              \if N\lasttitle \penalty\subsectionpenalty \fi
              \global\subsectionpenalty=-100
              \global\subsubsectionpenalty=10007
              \ifx#1*\noindent#2\else\stepCN{Tb}
              \setbox0=\hbox{\noindent\ArabicCN{Ta}.\ArabicCN{Tb}.}
              \indsect=\wd0\advance\indsect by 1em
              \parshape=2 0pt\hsize \indsect\dimensect
              \noindent\hbox to \indsect{\ArabicCN{Ta}.\ArabicCN{Tb}.\hfil}#1\fi
              \vskip\aftersubsect
              \egroup\let\lasttitle=Y
              \nobreak\parindent=0pt
              \everypar={\parindent=4mm
                         \penalty0\let\lasttitle=N}\ignorespaces}
\def\subsubsection#1#2 {\par\resetCN{Td}%
              \if N\lasttitle\else\vskip-\beforesubsubsect\fi
              \bgroup\tenpoint\sl
                 \setbox0=\vbox{\vskip\beforesubsubsect\noindent
              {\ArabicCN{Ta}.\ArabicCN{Tb}.\ArabicCN{Tc}.\kern1em}#1#2
              \vskip\aftersubsubsect}
              \dimen0=\ht0\advance\dimen0 by\dp0\advance\dimen0 by
                 2\baselineskip
              \advance\dimen0 by\pagetotal
              \ifdim\dimen0>\pagegoal
                 \ifdim\pagetotal>\pagegoal
                 \else \if N\lasttitle\eject\fi \fi\fi
              \vskip\beforesubsubsect
              \if N\lasttitle \penalty\subsubsectionpenalty \fi
              \global\subsubsectionpenalty=-50
              \ifx#1*\noindent#2\else\stepCN{Tc}
              \setbox0=\hbox{\noindent
              	\ArabicCN{Ta}.\ArabicCN{Tb}.\ArabicCN{Tc}.}
              \indsect=\wd0\advance\indsect by 1em
              \parshape=2 0pt\hsize \indsect\dimensect
              \noindent\hbox to 
              \indsect{\ArabicCN{Ta}.\ArabicCN{Tb}.\ArabicCN{Tc}.\hfil}#1\/\fi
              \vskip\aftersubsubsect
              \egroup\let\lasttitle=Y
              \nobreak\parindent=0pt
              \everypar={\parindent=4mm
                         \penalty0\let\lasttitle=N}\ignorespaces}
\def\paragraph#1#2 {\par\if N\lasttitle\else\vskip-\aftersubsubsect\fi
    		    \bgroup\tenpoint\rm
         	    \setbox0=\vbox{\vskip\aftersubsubsect\noindent
         	    {\ArabicCN{Ta}.\ArabicCN{Tb}.\ArabicCN{Tc}.\ArabicCN{Td}}#1#2}
              \dimen0=\ht0\advance\dimen0 by\dp0\advance\dimen0 by
                 2\baselineskip
              \advance\dimen0 by\pagetotal
              \ifdim\dimen0>\pagegoal
              \ifdim\pagetotal>\pagegoal
              \else \if N\lasttitle\eject\fi \fi\fi
              \vskip\aftersubsubsect
              \if N\lasttitle \penalty-50 \fi
              \ifx#1*\noindent\ul{#2:}\ \else\stepCN{Td}
              \setbox0=\hbox{\noindent
              	\ArabicCN{Ta}.\ArabicCN{Tb}.\ArabicCN{Tc}.\ArabicCN{Td}.}
              \indsect=\wd0\advance\indsect by 1em
              \parshape=2 0pt\hsize \indsect\dimensect
              \noindent\hbox to 
              \indsect{\ArabicCN{Ta}.\ArabicCN{Tb}.\ArabicCN{Tc}.\ArabicCN{Td}.
              \hfil}\ul{#1:}\ \fi
              \egroup\let\lasttitle=N}
%
\newtoks\ACK \ACK={Acknowledgements}
\def\ack#1{\par\vskip\beforesect\goodbreak
\noindent{\tenpoint\bf\the\ACK }\par\vskip\aftersect\penalty500\noindent#1}
%
\newtoks\APPND \APPND={Appendix}
\def\appendix#1{\par\vskip\beforesect\goodbreak
\noindent{\tenpoint\bf\the\APPND \kern.5em#1}\par
\vskip\aftersect\penalty500\let\lasttitle=Y}
\def\titleapp #1{\if N\lasttitle\goodbreak
\else\vskip-\beforesubsect\penalty500\fi
\vskip\beforesubsect
{\tenpoint\bf\noindent\ignorespaces #1}
\vskip\aftersubsect\let\lasttitle=Y\noindent}
%
\newtoks\REFNAME \REFNAME={References}
\def\references{\begREF}
\def\begREF{\bgroup
              \setbox0=\vbox{\vskip\beforesect\noindent{\bf\the\REFNAME}
                       \vskip\aftersect}
              \dimen0=\ht0\advance\dimen0 by\dp0 
              \advance\dimen0 by 2\baselineskip
              \advance\dimen0 by\pagetotal
              \ifdim\dimen0>\pagegoal
                 \ifdim\pagetotal>\pagegoal
                 \else\eject\fi\fi
              \vskip\beforesect\noindent{\bf\the\REFNAME}
              \vskip\aftersect
               \frenchspacing \parindent=0pt \leftskip=1truecm
               \everypar{\hangindent=\parindent}}
\def\ref#1{ $[{\rm #1}]$}%
\gdef\refis#1{\item{$[$#1$]$\ }}   
\def\endreferences{\par\egroup}
%
\def\review#1, #2, 1#3#4#5, #6 {{\sl#1\/} {\bf#2} (1#3#4#5) #6}
\def\book#1, #2, #3, 1#4#5#6, #7 {#1 (#2, #3, 1#4#5#6) p. #7}
%
\newcount\foCN \foCN=0
\def\fonote{\global\advance\foCN by 1
$(^{\rm\number\foCN})$\vfootnote{$(^{\rm\number\foCN})$}}
%
\catcode`@=11
\def\vfootnote#1{\insert\footins\bgroup
  \ninepoint
  \interlinepenalty\interfootnotelinepenalty
  \splittopskip\ht\strutbox 
  \splitmaxdepth\dp\strutbox \floatingpenalty\@MM
  \leftskip\z@skip \rightskip\z@skip \spaceskip\z@skip \xspaceskip\z@skip
  \baselineskip=10pt\lineskip=10pt
  \noindent\kern10mm\llap{#1\enspace}\footstrut
  \futurelet\next\fo@t}
\def\@foot{\egroup}
\catcode`@=12
%
\newtoks\shorttitle
\newtoks\authors
\shorttitle={SHORT TITLE}%
\authors={The Authors}%
\def\firsthd{\hfill}
\def\hdleft{\tenpoint\folio\hfill{\sl \the\authors\/}\hfill}
\def\hdrigt{\tenpoint\hfill{\ninepoint\the\shorttitle}\hfill\folio}
\newif\ifbegpage
\headline={\ifbegpage\firsthd
\global\begpagefalse\else
\ifodd\pageno\hdrigt\else\hdleft\fi
\advance\pageno by 1\fi}
\footline={\hfil}
\def\makeheadline{\vbox to0pt{\vskip-10mm
\line{\vbox to8.5pt{}\the\headline}\vss}\nointerlineskip}
%
\begpagetrue
%
\newtoks\TABLE \TABLE={Table}
%
\newdimen\tableheight
\newskip\superskipamount \superskipamount=28pt plus 4pt minus 4pt
\def\superskip{\vskip\superskipamount}
\def\superbreak{\par\ifdim\lastskip<\superskipamount
  \removelastskip\penalty-200\superskip\fi}
\catcode`\@=11
\def\endinsert{\egroup 
  \if@mid \dimen@\ht\z@ \advance\dimen@\dp\z@
    \advance\dimen@24\p@ \advance\dimen@\pagetotal
    \ifdim\dimen@>\pagegoal\@midfalse\p@gefalse\fi\fi
  \if@mid \superskip\box\z@\superbreak
  \else\insert\topins{\penalty100 
    \splittopskip\z@skip
    \splitmaxdepth\maxdimen \floatingpenalty\z@
    \ifp@ge \dimen@\dp\z@
    \vbox to\vsize{\unvbox\z@\kern-\dimen@}
    \else \box\z@\nobreak\superskip\fi}\fi\endgroup}
\catcode`\@=12
\newcount\CNfig \CNfig=0
\let\captext=N
\def\begfig #1cm{\midinsert\tableheight=#1cm\advance\tableheight by 5mm
	\vglue\tableheight}
\def\topfig #1cm{\topinsert\tableheight=#1cm\advance\tableheight by 5mm
	\vglue\tableheight}
\long\def\caption#1{\let\captext=Y\stepCN{fig}\ninepoint\rm
	\noindent Fig.\ \number\CNfig.\kern.3em ---\kern.3em\ignorespaces
     \parindent=0pt#1\par}
\def\endfig{\if N\captext\stepCN{fig}
	\ninepoint\rm\noindent Figure \number\CNfig\else\let\figtext=N\fi
	\endinsert}
%
\newcount\CNtab \CNtab=0
\let\tabtext=N
\def\begtab #1cm{\midinsert\tableheight=#1cm}
\def\toptab #1cm{\topinsert\tableheight=#1cm}
\long\def\tabcap#1{\let\tabtext=Y\stepCN{tab}
	\ninepoint\rm\noindent Table\ \RomanCN{tab}.\kern.3em ---\kern.3em\ignorespaces
	\parindent=0pt#1\par}
\def\endtab{\if N\tabtext\stepCN{tab}
	\ninepoint\rm\noindent Table \RomanCN{tab} \else\let\tabtext=N\fi
	\ifdim\tableheight=0cm\vskip-\belowdisplayskip
     \else\advance\tableheight by 5mm\fi
     \vglue\tableheight\endinsert}
%

%
\def\(#1){(\call{#1})}

\catcode`@=11
\newcount\r@fcount \r@fcount=0
\newcount\r@fcurr
\immediate\newwrite\reffile
\newif\ifr@ffile\r@ffilefalse
\def\w@rnwrite#1{\ifr@ffile\immediate\write\reffile{#1}\fi\message{#1}}
 
\def\writer@f#1>>{}
\def\referencefile{
  \r@ffiletrue\immediate\openout\reffile=\jobname.ref%
  \def\writer@f##1>>{\ifr@ffile\immediate\write\reffile%
    {\noexpand\refis{##1} = \csname r@fnum##1\endcsname = %
     \expandafter\expandafter\expandafter\strip@t\expandafter%
     \meaning\csname r@ftext\csname r@fnum##1\endcsname\endcsname}\fi}%
  \def\strip@t##1>>{}}

\def\citeall#1{\xdef#1##1{#1{\noexpand\cite{##1}}}}
\def\cite#1{\each@rg\citer@nge{#1}}
 
\def\each@rg#1#2{{\let\thecsname=#1\expandafter\first@rg#2,\end,}}
\def\first@rg#1,{\thecsname{#1}\apply@rg}
\def\apply@rg#1,{\ifx\end#1\let\next=\relax
\else,\thecsname{#1}\let\next=\apply@rg\fi\next}
 
\def\citer@nge#1{\citedor@nge#1-\end-}
\def\citer@ngeat#1\end-{#1}
\def\citedor@nge#1-#2-{\ifx\end#2\r@featspace#1 
  \else\citel@@p{#1}{#2}\citer@ngeat\fi}
\def\citel@@p#1#2{\ifnum#1>#2{\errmessage{Reference range #1-#2\space is bad.}%
    \errhelp{If you cite a series of references by the notation M-N, then M and
    N must be integers, and N must be greater than or equal to M.}}\else%
 {\count0=#1\count1=#2\advance\count1 by1\relax\expandafter\r@fcite\the\count0,%
  \loop\advance\count0 by1\relax
    \ifnum\count0<\count1,\expandafter\r@fcite\the\count0,%
  \repeat}\fi}
 
\def\r@featspace#1#2 {\r@fcite#1#2,}
\def\r@fcite#1,{\ifuncit@d{#1}
    \newr@f{#1}%
    \expandafter\gdef\csname r@ftext\number\r@fcount\endcsname%
                     {\message{Reference #1 to be supplied.}%
                      \writer@f#1>>#1 to be supplied.\par}%
 \fi%
 \csname r@fnum#1\endcsname}
\def\ifuncit@d#1{\expandafter\ifx\csname r@fnum#1\endcsname\relax}%
\def\newr@f#1{\global\advance\r@fcount by1%
    \expandafter\xdef\csname r@fnum#1\endcsname{\number\r@fcount}}
 
\let\r@fis=\refis
\def\refis#1#2#3\par{\ifuncit@d{#1}
   \newr@f{#1}%
   \w@rnwrite{Reference #1=\number\r@fcount\space is not cited up to now.}\fi%
  \expandafter\gdef\csname r@ftext\csname r@fnum#1\endcsname\endcsname%
  {\writer@f#1>>#2#3\par}}
 
\def\ignoreuncited{
   \def\refis##1##2##3\par{\ifuncit@d{##1}%
     \else\expandafter\gdef\csname r@ftext\csname r@fnum##1\endcsname\endcsname%
     {\writer@f##1>>##2##3\par}\fi}}
 
\def\r@ferr{\endreferences\errmessage{I was expecting to see
\noexpand\endreferences before now;  I have inserted it here.}}
\let\r@ferences=\references
\def\references{\r@ferences\def\endmode{\r@ferr\par\endgroup}}
 
\let\endr@ferences=\endreferences
\def\endreferences{\r@fcurr=0
  {\loop\ifnum\r@fcurr<\r@fcount
    \advance\r@fcurr by 1\relax\expandafter\r@fis\expandafter{\number\r@fcurr}%
    \csname r@ftext\number\r@fcurr\endcsname%
  \repeat}\gdef\r@ferr{}\endr@ferences}
 
 
\let\r@fend=\endpaper\gdef\endpaper{\ifr@ffile
\immediate\write16{Cross References written on []\jobname.REF.}\fi\r@fend}
 
\catcode`@=12

\citeall\ref%

\catcode`@=11
\newcount\tagnumber\tagnumber=0
 
\immediate\newwrite\eqnfile
\newif\if@qnfile\@qnfilefalse
\def\write@qn#1{}
\def\writenew@qn#1{}
\def\w@rnwrite#1{\write@qn{#1}\message{#1}}
\def\@rrwrite#1{\write@qn{#1}\errmessage{#1}}
 
\def\taghead#1{\gdef\t@ghead{#1}\global\tagnumber=0}
\def\t@ghead{}
 
\expandafter\def\csname @qnnum-3\endcsname
  {{\t@ghead\advance\tagnumber by -3\relax\number\tagnumber}}
\expandafter\def\csname @qnnum-2\endcsname
  {{\t@ghead\advance\tagnumber by -2\relax\number\tagnumber}}
\expandafter\def\csname @qnnum-1\endcsname
  {{\t@ghead\advance\tagnumber by -1\relax\number\tagnumber}}
\expandafter\def\csname @qnnum0\endcsname
  {\t@ghead\number\tagnumber}
\expandafter\def\csname @qnnum+1\endcsname
  {{\t@ghead\advance\tagnumber by 1\relax\number\tagnumber}}
\expandafter\def\csname @qnnum+2\endcsname
  {{\t@ghead\advance\tagnumber by 2\relax\number\tagnumber}}
\expandafter\def\csname @qnnum+3\endcsname
  {{\t@ghead\advance\tagnumber by 3\relax\number\tagnumber}}
 
\def\equationfile{%
  \@qnfiletrue\immediate\openout\eqnfile=\jobname.eqn%
  \def\write@qn##1{\if@qnfile\immediate\write\eqnfile{##1}\fi}
  \def\writenew@qn##1{\if@qnfile\immediate\write\eqnfile
    {\noexpand\tag{##1} = (\t@ghead\number\tagnumber)}\fi}
}
 
\def\callall#1{\xdef#1##1{#1{\noexpand\call{##1}}}}
\def\call#1{\each@rg\callr@nge{#1}}
 
\def\each@rg#1#2{{\let\thecsname=#1\expandafter\first@rg#2,\end,}}
\def\first@rg#1,{\thecsname{#1}\apply@rg}
\def\apply@rg#1,{\ifx\end#1\let\next=\relax%
\else,\thecsname{#1}\let\next=\apply@rg\fi\next}
 
\def\callr@nge#1{\calldor@nge#1-\end-}
\def\callr@ngeat#1\end-{#1}
\def\calldor@nge#1-#2-{\ifx\end#2\@qneatspace#1 %
  \else\calll@@p{#1}{#2}\callr@ngeat\fi}
\def\calll@@p#1#2{\ifnum#1>#2{\@rrwrite{Equation range #1-#2\space is bad.}
\errhelp{If you call a series of equations by the notation M-N, then M and
N must be integers, and N must be greater than or equal to M.}}\else%
 {\count0=#1\count1=#2\advance\count1 by1\relax\expandafter\@qncall\the\count0,%
  \loop\advance\count0 by1\relax%
    \ifnum\count0<\count1,\expandafter\@qncall\the\count0,%
  \repeat}\fi}
 
\def\@qneatspace#1#2 {\@qncall#1#2,}
\def\@qncall#1,{\ifunc@lled{#1}{\def\next{#1}\ifx\next\empty\else
  \w@rnwrite{Equation number \noexpand\(>>#1<<) has not been defined yet.}
  >>#1<<\fi}\else\csname @qnnum#1\endcsname\fi}
 
\let\eqnono=\eqno
\def\eqno(#1){\tag#1}
\def\tag#1$${\eqnono(\displayt@g#1 )$$}
 
\def\aligntag#1\endaligntag
  $${\gdef\tag##1\\{&(##1 )\cr}\eqalignno{#1\\}$$
  \gdef\tag##1$${\eqnono(\displayt@g##1 )$$}}

\def\eqalignno#1{\displ@y \tabskip\centering
  \halign to\displaywidth{\hfil$\displaystyle{##}$\tabskip\z@skip
    &$\displaystyle{{}##}$\hfil\tabskip\centering
    &\llap{$\displayt@gpar##$}\tabskip\z@skip\crcr
    #1\crcr}}
 
\def\displayt@gpar(#1){(\displayt@g#1 )}
 
\def\displayt@g#1 {\rm\ifunc@lled{#1}\global\advance\tagnumber by1
        {\def\next{#1}\ifx\next\empty\else\expandafter
        \xdef\csname @qnnum#1\endcsname{\t@ghead\number\tagnumber}\fi}%
  \writenew@qn{#1}\t@ghead\number\tagnumber\else
        {\edef\next{\t@ghead\number\tagnumber}%
        \expandafter\ifx\csname @qnnum#1\endcsname\next\else
        \w@rnwrite{Equation \noexpand\tag{#1} is a duplicate number.}\fi}%
  \csname @qnnum#1\endcsname\fi}
 
\def\ifunc@lled#1{\expandafter\ifx\csname @qnnum#1\endcsname\relax}
 
\let\@qnend=\end\gdef\end{\if@qnfile
\immediate\write16{Equation numbers written on []\jobname.EQN.}\fi\@qnend}
 
\catcode`@=12
 
 

\input epsf
\input psfig.sty

\def \d {\delta x}

\def \ov  {\overline}

\authors{Rama CONT, Marc POTTERS \& Jean-Philippe BOUCHAUD}
\shorttitle{Scaling in financial markets}

%
\CNlecture=14
%
\title{Scaling in stock market data:\\
stable laws and beyond}
\author{Rama CONT\inst{1,2,3},
Marc POTTERS\inst{2} and Jean-Philippe BOUCHAUD\inst{1,2}}

\institute{\inst{1} Service de Physique de l'Etat Condens\'e\\
Centre d'Etudes de Saclay 91191 Gif-sur-Yvette, FRANCE\\
\inst{2} Science \& Finance\\
109-111 rue Victor Hugo\\
92532 Levallois, FRANCE\\
\inst{3}  Laboratoire de Physique de la Mati\`ere Condens\'ee\\
CNRS URA 190, Universit\'e de Nice \\06108 Nice Cedex 2, France
}

\maketitle

\section{Introduction}

The concepts of scale invariance and scaling behavior are now increasingly
 applied outside their traditional domains of application, the physical
sciences. Their application to financial markets, initiated by Mandelbrot \ref{mandelbrot1,mandelbrot2} in the 1960s, has experienced a regain of interest in the recent years, partly due to the abundance of high-frequency data sets and
availability of computers for analyzing their statistical properties.

This lecture is intended as an introduction and a brief review of
current research in a field which is becoming increasingly popular in the
theoretical physics community. We will try to show how the concepts of scale
invariance and scaling behavior may be usefully applied in the framework
of a statistical approach to the study of financial data, pointing out at the 
same time the limits of such an approach. 

\section{Statistical description of market data}

Quantitative research mainly focuses on "liquid" financial markets i.e. organized 
markets where
transactions are frequent and the number of actors is large. Typical examples
 are foreign exchange markets, organized futures markets and stock index
markets and the market for large stocks. Prices are recorded several times a minute in such markets, creating
mines of data to exploit.
Such markets are complex systems with many degrees of freedom\ref{santafe}, where many internal and external factors
interact at each instant in order to fix the transaction price of financial assets.Various  factors
 such as public policy,  interest rates and economic conditions doubtlessly
influence market behavior. However, the precise nature of their influence
is not well known and  given the complex nature of
the pricing mechanism, simple deterministic models are unable to reproduce
the properties observed in financial time series. Furthermore, although the details of the price fixing mechanism- market microstructure - may be different
from one market to the other, what is striking is the universality of some
simple statistical properties of price fluctuations, prompting a unified approach to the study of different types of markets.

As in the case of other types of complex systems with universal characteristics,
 a stochastic approach has proved to
be fruitful in this case. The main object of study in this framework is the probability density function
(PDF) of the increments at a given time scale $T$ i.e.
 the probability distribution  $P_T$ of $x(t+T) - x(t)$ where $x(t)$ is the price of the asset at time $t$. This approach was inaugurated by Louis Bachelier who
first introduced the idea that stock market prices behave as a random walk \ref{bachelier},
and who considered Brownian motion  as a candidate for modeling price fluctuations.

Bachelier's model, applied to the logarithm of the prices (to ensure positivity of
the price!), became very popular in the 1950s \ref{cootner} and it is one of the main
ingredients of the famous Black-Scholes option pricing formula \ref{bs}.
This model implies that the increments of asset returns (or asset prices)
are independent identically distributed (iid) Gaussian variables.
Indeed if one considers each price change as   a sum of
many small and independent random contributions from various market factors,
the Central Limit Theorem suggests the Gaussian as a natural candidate.

Figure 1. gives a typical example of the empirical distribution of the 
increments of asset prices, in this case the U.S. Dollar/ Deutschemark exchange rate,
sampled every 5 minutes. The contrast with the Gaussian is striking: this 'heavy tailed'
or 'leptokurtic' character of the distribution of price changes, has been
repeatedely observed in various market data and may be quantitavely measure by the kurtosis of the distribution $P_T$ defined by :

$$
 \kappa = {{\overline{(\Delta_T x - \overline{\Delta_T x})^4}}\over {\sigma(T) ^ 4 } } - 3 \tag(kurtosis)
$$
where $\sigma(T)$ is the variance of $\Delta_T x =x(t+T)-x(t)$.
The kurtosis is defined such that $\kappa = 0$ for a Gaussian distribution, a positive value
of $\kappa$ indicating a slow asymptotic decay of the PDF. The kurtosis
of the increments of asset prices is far from its Gaussian value: typical values for T=5 minutes are: $\kappa \simeq 74$ (US\$/DM exchange rate futures),
$\kappa \simeq 60$ (US\$/Swiss Franc exchange rate futures), 
$\kappa \simeq 16$ (S\&P500 index futures) \ref{cont,book}.

These observations imply that using a Gaussian PDF systematically underestimates the probability
of large price fluctuations, an issue of utmost importance is financial risk management.
 
\psfig{file=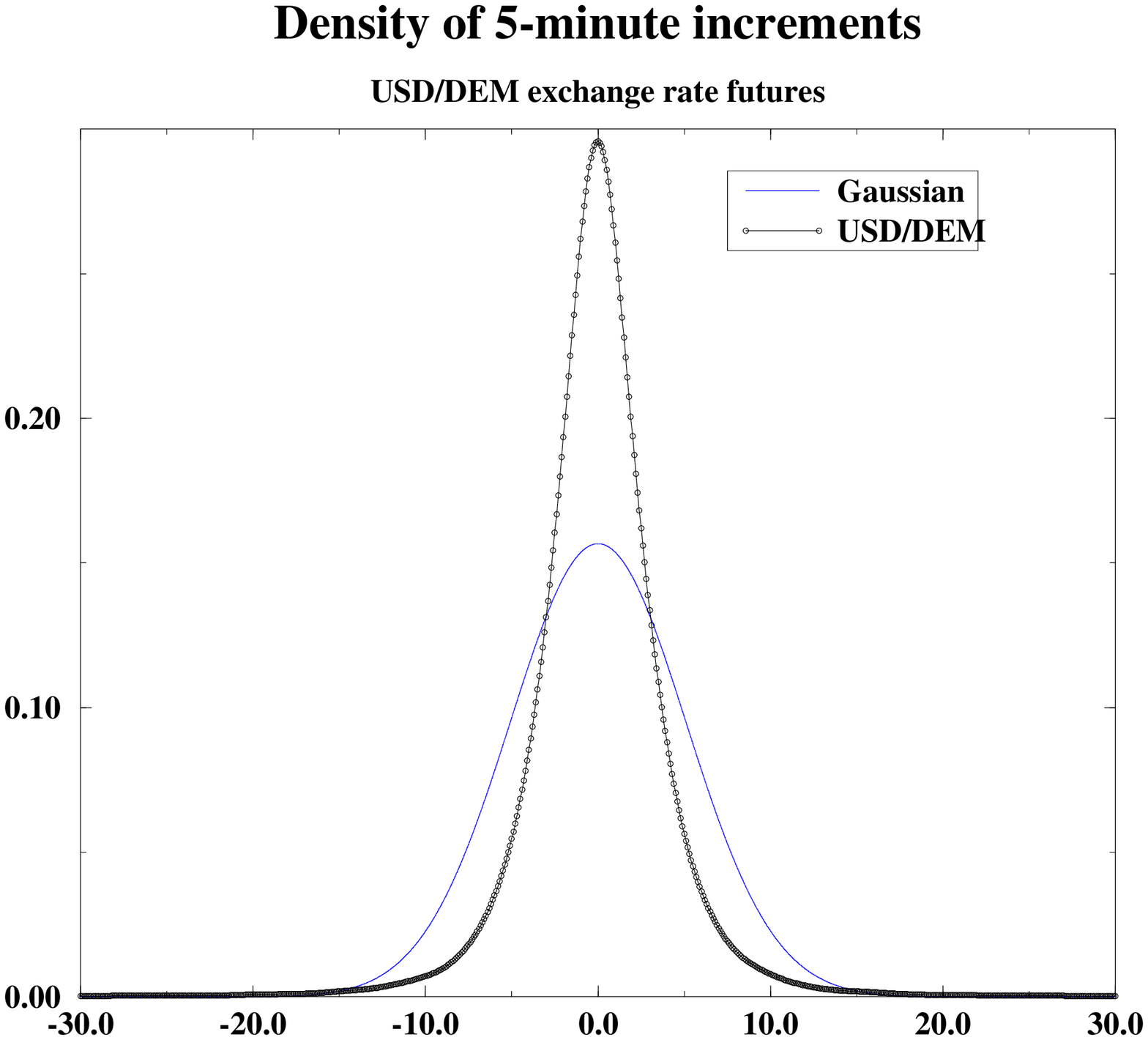,width=10cm,angle=270}
\caption{ Probability density of 5 minute increments of USD/Deutschemark exchange rate futures. The lower curve is a gaussian with same mean and variance.}
 
\section{Scale invariance and stable laws}

But why should stock price fluctuations have scale invariant properties in the first place?
The answer lies in a generalized version of the Central Limit Theorem \ref{gnedenko}: the  distribution of the sum a large number of independent identically distributed random variables  belongs to a family of distributions known as {\it stable} or {\it stable L\'evy} distributions, characterized by
their Fourier transform (characteristic function) \ref{gnedenko,bouchaud}:

$$
\phi_\mu(z) = \exp{[-a_\mu  |z|^{\mu} (1 + i\beta \tan{(\mu\pi/2)} {{z}\over{|z|}})]} \tag{levy}
$$
The parameter $\mu$, called the stability index, belongs to the interval ]0,2], $\mu=2$
corresponding to the Gaussian distribution. $\beta$ is a skewness parameter, $\beta=0$ for a symmetric distribution. L\'evy distributions are characterized by the property of being stable under convolution: in other words, the sum of two
{\it iid} L\'evy-distributed variables is also L\'evy distributed, with the same stability index $\mu$. More precisely, if $X_i, i=1..n$ are n {\it iid} $\mu$-stable random variables
then the renormalized sum:
$$
S_n = {{\sum_{i=1}^{n}X_i}\over{N^{1/\mu}}}
$$
has the same PDF as the $X_i$.
In particular, the sum scales as $N^{1/\mu}$ and not as $\sqrt{N}$ which is the case
in diffusive random walks. 
L\'evy distributions, in addition
to their stability under convolution, have other interesting properties:
except for the Gaussian ($\mu=2$) all $\mu$-stable PDFs have power-law tails
with exponent $1+\mu$:

$$p_{\mu}(x) \simeq {{C}\over{x^{1+\mu}}} \tag{decay}$$

leading to an infinite variance and heavy tails.

These observations led Mandelbrot to propose stable
L\'evy  distributions as candidates for the PDF of price changes of financial assets \ref{mandelbrot1,mandelbrot2}. Mandelbrot observed that stable distributions  offer a heavy tailed alternative to Bachelier's model while
still enabling a justification of the model in terms of the Central Limit Theorem (see above). Furthermore, their stability under convolution gives rise
to the scale invariance of the process: if appropriately rescaled, the increment
at scale $N\tau$ will have the same distribution as the increment at scale $\tau$:

$$
P_{N\tau}(x) ={{1}\over{\lambda}} P_\tau({{x}\over{\lambda}}) \ , \  \ \lambda = N^{1/\mu}
$$

The above relation means that the price process $x(t)$ is {\it self-similar}  with a self-similarity exponent which is   the inverse of the stability index $\mu$. Self-similarity in market prices was first remarked by Mandelbrot in 
the 1960s in his seminal work cotton prices \ref{mandelbrot1}. 
Recent studies have confirmed the presence of self-similarity and scale invariant properties in various markets:  the Milan stock exchange \ref{mantegna1}, the S\&P500 index \ref{mantegna2}, the CAC40 (Paris stock market) index \ref{zaj} and foreign exchange markets \ref{olsen,olsen2} and individual french stocks \ref{belkacem}. The value of $\mu$ may be estimated
for example by examining the scaling behavior of the probability of return to the origin $P_T(0)$, which scales as ${{1}\over N^{1/\mu}}$, which is indeed
the method used in \ref{mantegna2,zaj}. The value of $\mu$ found depends on the market considered but all
values fall between 1.5 and 1.7, $\mu$ being lower for more volatile markets. 

\midinsert
\vskip 0.5cm
\centerline{ 
\psfig{file=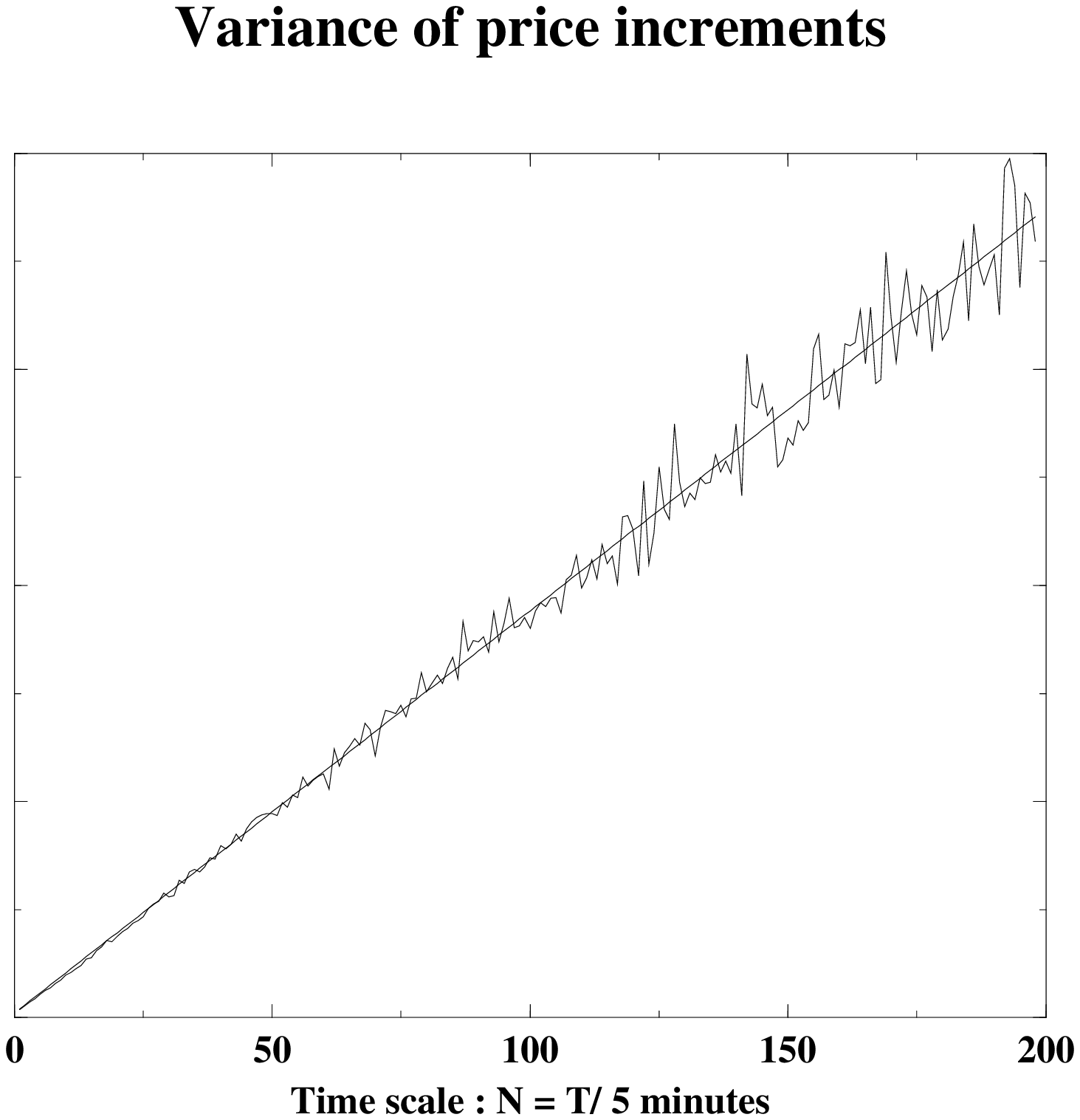,width=5cm,angle=0}
\hskip 0.9cm
\psfig{file=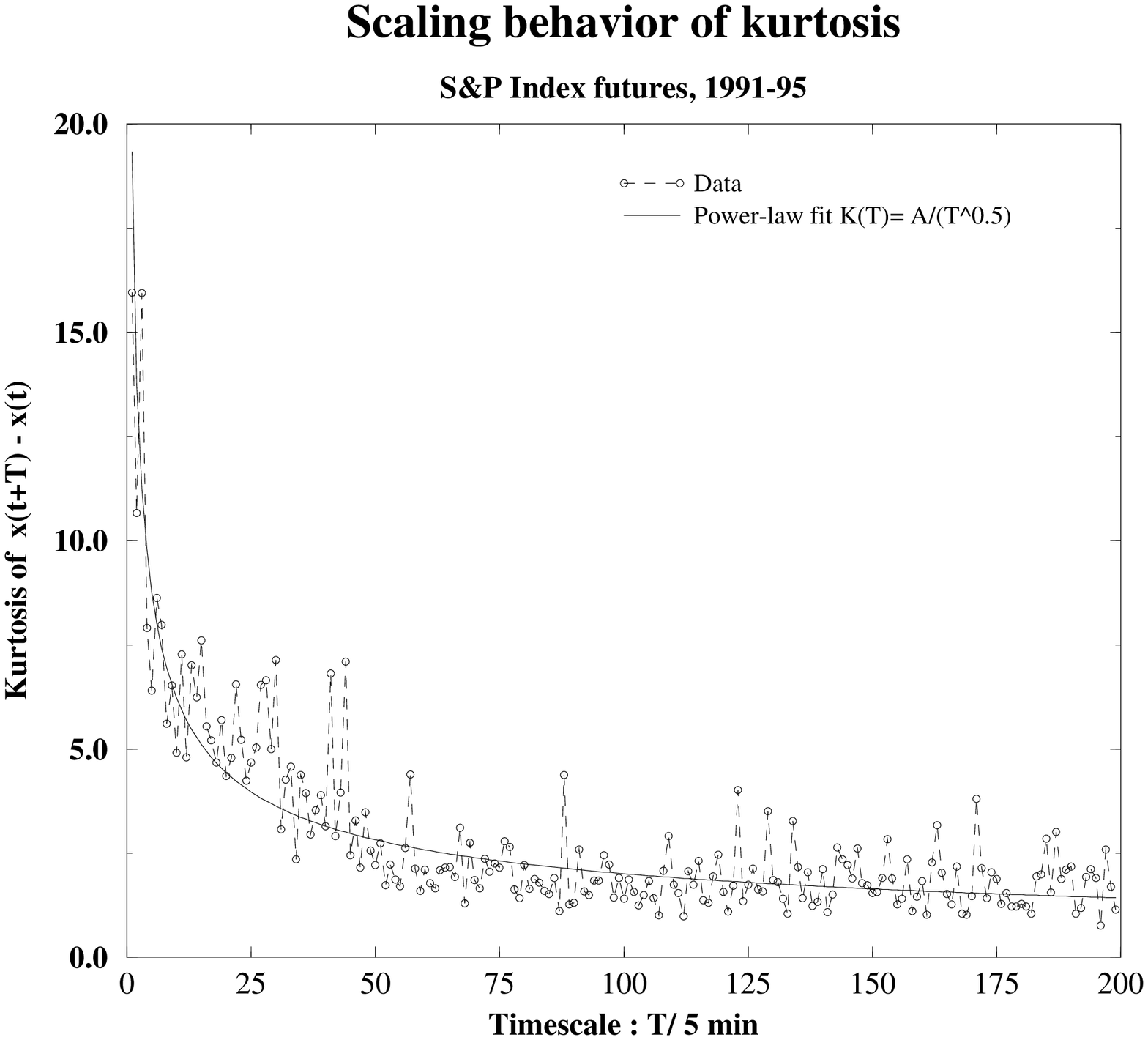,width=5cm,angle=270}}
\vskip 0.5cm
\caption{Left:~Scaling behavior of the variance of price increments for 
S\&P 500 index futures: the variance increases linearly with the time scale,
a property which is consistent with the absence of significant linear correlation.
Right:~Scaling behavior of the kurtosis of price increments for S\&P 500 index futures: the kurtosis decreases more slowly than in the case of iid increments where it decreases as $1/N$. The solid line represents a fit with $\kappa(T) = \kappa(1)/\sqrt{T}$.
}
\endinsert

\section{Beyond scale invariance: truncated L\'evy flights}

Although the representation of price increments as L\'evy-stable {\it iid} variables accounts for their leptokurtic character as well as their self-similar properties, it has several drawbacks:
careful comparison of the distributions at various time scales show that
in fact it is {\it not} scale invariant \ref{book,olsen2,cont}. The self-similarity
properties described above do not hold at all time scales but only for short ones, typically less than a week. For example, if the price followed
a L\'evy walk, different methods of estimating the self-similarity parameter
would give the same value, namely $1/\mu$. Yet in practice, data from different
time resolutions give different  values of the self-similarity exponent,
between 0.5 (the value for a Brownian diffusion) and 0.6 ($=1/\mu$ with $\mu= 1.7$).
 
Furthermore, the best fit by a stable L\'evy distribution \ref{mantegna2}
still overestimates the probability for extreme fluctuations: in other words,
L\'evy distributions have tails that are 'too fat' compared to fluctuations of real prices. Tails of empirical distributions may be studied more closely by  rank-ordering techniques \ref{book,these} (also called Zipf plots). They turn out to decay more slowly than a Gaussian but more quickly than any L\'evy-stable distribution: their decay is better described by an exponential tail. Last but not least, real price changes turn out to have finite variance \ref{book,olsen2}.

These apparently contradictory aspects- self-similarity  at short time scales, breakdown of scaling for longer time scales, truncated power law tails, finite variance- may be blended together into a consistent picture by 
  a {\it truncated L\'evy flight} description\ref{mantegna3,book,comment}. Consider
a random walk with increments distributed with an exponentially truncated L\'evy density $p(x)$ \ref{koponen}. $p(x)$ may be characterized by its Fourier transform (characteristic function):
 
$$\phi(z) = \exp{(-a_\mu[((\alpha^2 + z^2)^{\mu/2}{{\cos{(\mu\ arctan(|z|/\alpha) )}}\over{\cos{(\pi\mu/2)}}}-\alpha^\mu)]   )} \tag{tlf}$$ 

The above expression differs from the characteristic function of a stable law
by the presence of a cut-off parameter $\alpha$.
$\phi(z)$ is defined such that it has a regular Taylor expansion at the origin,
assuring the finiteness of all moments. In terms of the variable $x$, $p(x)$ decays exponentially for large values of $x$, which describes adequately the decay of the tails of the PDF of empirical price changes. The asymptotic 
behavior of $\phi$ is given by

$$\phi(z)  \mathop{\sim}_{z\to\infty} \phi_{\mu}(z)$$
 i.e. for small values
of $x$, $p(x)$ behaves like a L\'evy-stable law of index $\mu$. We can now proceed to describe
the behavior of a random walk with {\it iid} increments having a truncated L\'evy PDF $p(x)$. For short time scales, $p(x)$ behaves like a stable law of
index $\mu$: the short term dynamics of the process is therefore correctly
described by a L\'evy flight which gives rise to the scale invariance with a self-similarity exponent $H = 1/\mu$
\ref{mandelbrot1,mandelbrot2,mantegna1,mantegna2}. However, since  the
variance of the distribution is finite, the  Gaussian version of the Central Limit Theorem applies   and for large time scales $T \gg T^* = \kappa\tau$, where $\kappa$
is the kurtosis of the distribution $P_\tau(x)$. Thus at large time resolutions
the distribution $P_T$ will approach a Gaussian, which is consistent
with the fact that the kurtosis tends to zero when $T\to\infty$ (see figure). 
Between these two regimes -L\'evy flight for short times and Gaussian diffusion for long times- is a crossover regime which is characterized by the appearance of fluctuations of the order $1/\alpha$.

\midinsert
\vskip 0.5cm
\centerline{ 
\psfig{file=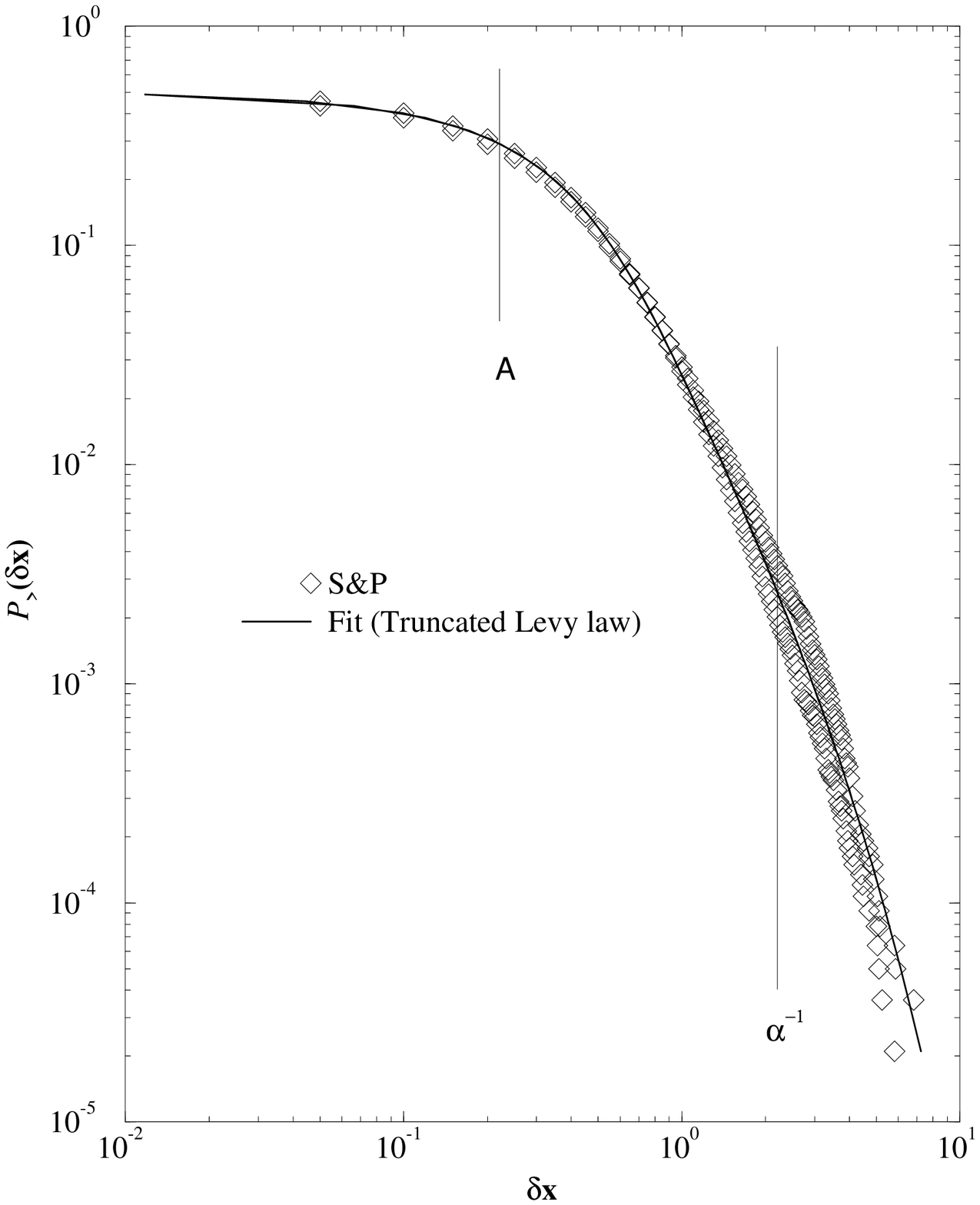,height=8cm,width=8cm,angle=0}}
\vskip 0.5cm
\caption{Rank ordering of the increments of S\&P 500 index futures. 
The cumulative distribution function of price increments
is well represented by an exponentially truncated L\'evy distribution with 
finite variance.}
\endinsert

\section{Correlation and dependence}

It is a well-known fact that price movements in liquid markets
do not exhibit any significant autocorrelation: the autocorrelation function of the price changes

$$ C(T) = {{\ov{ \d_{t}\d_{t+T}} -  \ov{ \d_{t}}\  \ov{ \d_{t+T}} }\over{var(\d_t)}}$$
  
rapidly decays to zero in a few minutes: for  $T\geq15$ minutes it can be safely  assumed to
be zero for all practical purposes\ref{comment}. The absence of significant
linear correlations in
price increments and asset returns has been widely documented \ref{fama1,pagan} and often cited
as support for the "Efficient Market Hypothesis" \ref{fama2}.
The absence of correlation is intuitively easy to understand:
if price changes exhibit significant correlation, this correlation may be used
to conceive a simple strategy with  positive expected earnings; such strategies,
termed {\it arbitrage}, will therefore tend to reduce correlations except for very short time scales, which represent the time the market takes to react to new information. This correlation time is typically several minutes for organized futures markets and even shorter for foreign exchange markets.

 The fast decay
of the correlation function implies the additivity of variances: for
uncorrelated variables, the variance of the sum is the sum of the variances.
The absence of linear correlation is thus consistent with the linear increase
of the variance with respect to time scale.

However, the absence of serial correlation does not imply the independence of the increments \ref{cont}:  for example the square or the absolute value of price changes exhibits 
slowly decaying serial correlations.
Figure 5 displays the autocorrelation function $g(T)$
of the square of the increments, defined as:

$$
g(T) =  {{\overline{\d_t^2 \d_{t+T}^2 } - \overline{\d_t^2}\ \overline{ \d_{t+T}^2 }}\over{var (\d ^2)}} =  
{{\overline{\d_t^2 \d_{t+T}^2 } - \overline{\d_t^2}\ \overline{ \d_{t+T}^2 }}\over{\mu_4(\tau) - \sigma(\tau)^4} } $$

for S\&P 500 Futures.The slow decay of $g(T)$ (see figure) is well  represented by a power law \ref{cont}:

$$
g(k) \simeq {{g_0}\over{k^\alpha}}\qquad \alpha = 0.37 \pm 0.037 \qquad g_0 = 0.08
$$
 
This slow relaxation of the correlation function $g$ indicated persistence in
the {\it scale} of fluctuations. Other measures of the scale of fluctuations,
such as the absolute value of the increments, exhibit the same type of persistence, a phenomenon  which can be related to the "clustering of volatility", well known in the financial literature: a large price movement
tends to be followed by another large price movement but not necessarily in the same direction.

\vskip 0.5cm
\centerline{ 
\psfig{file=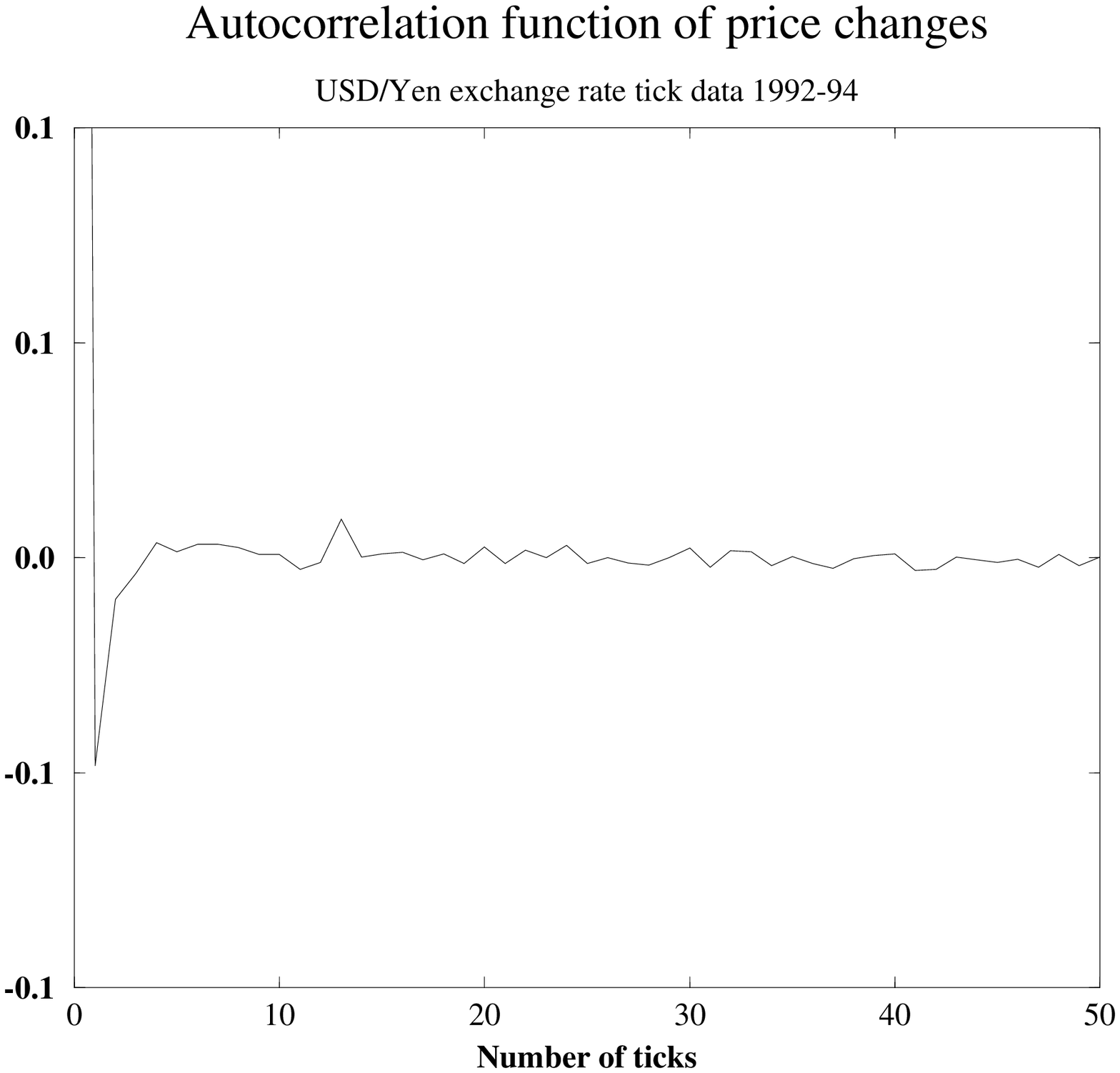,width=6cm,angle=270}
\psfig{file=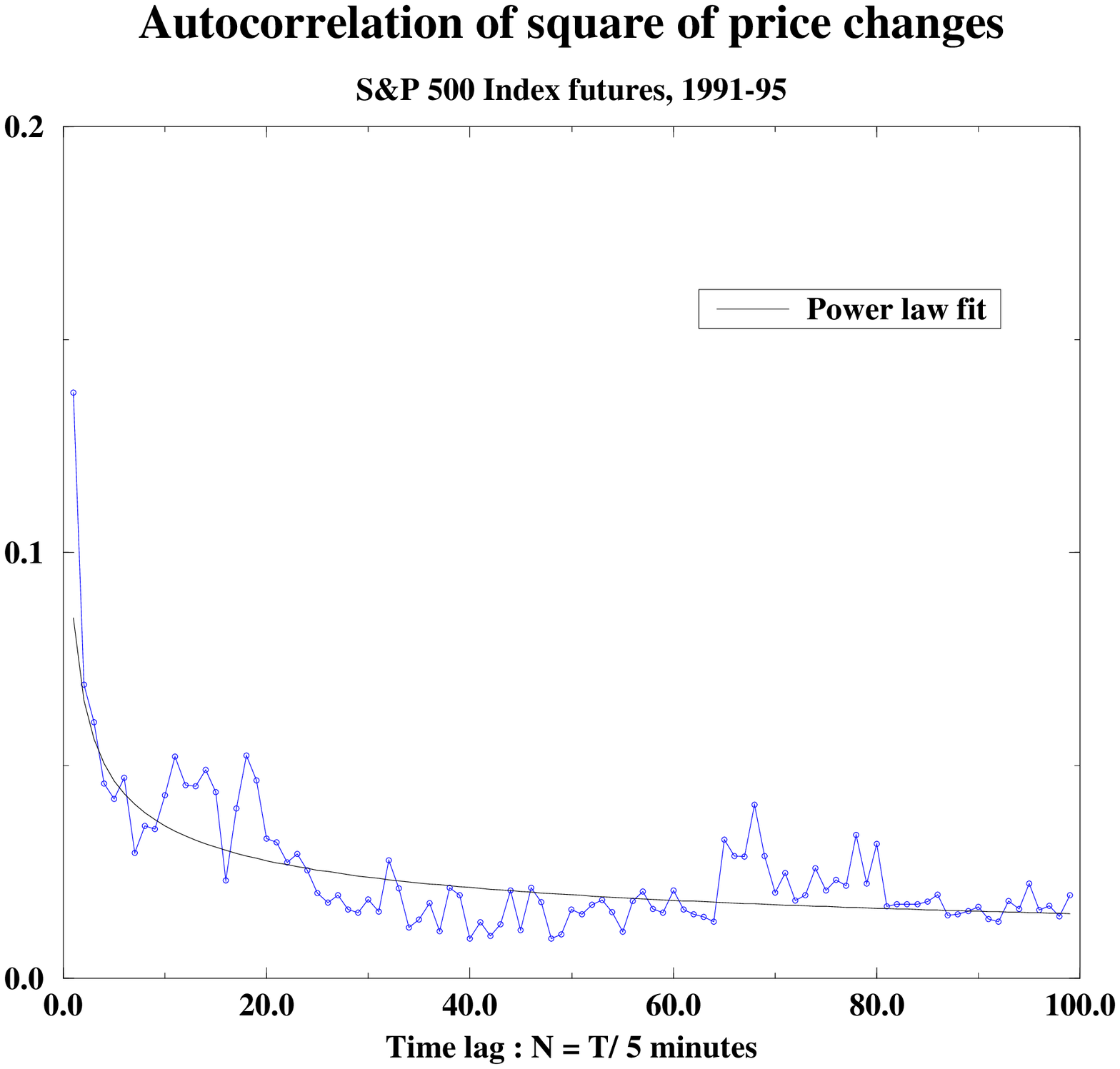,width=6cm,angle=270}}
\vskip 0.5cm
\caption{Left:~Autocorrelation function of price increments for the US Dollar/ Yen exchange rate.
Right:~The autocorrelation function g of square of price increments of S\&P futures exhibits
a slow decay, well represented by a power law with exponent $\alpha\simeq 0.37$.
}

The persistent character of the scale of fluctuations may be quantitatively
related to the anomalous scaling behavior i.e. slow decrease of the
kurtosis as the time resolution increases. As a result, the convergence
to the Gaussian is slowed down even more than in the case of a truncated L\'evy flight with {\it iid} increments.

The presence of such subtle correlations points out the limits of a representation of market prices by a random walk model and the need to take into account finer effects due to nonlinear correlations and non-stationarity \ref{adapt,cont}. These effects are especially important in options markets, where the behavior of volatility -the scale of price fluctuations- is crucial in determining the value of the option. It has been shown \ref{adapt} that the
correlations present in the volatility -as represented by the function $g$ defined above- are accurately reflected in market prices for
options.

\section{Turbulence and finance}

Recently some attempts have been made to draw analogies between
scaling properties of foreign exchange rates and turbulent flows 
\ref{peinke}, based on some similarities between their probability density functions. Indeed, the statistical approach to the problem is the same, price increments
playing the same role as velocity increments in the case of turbulence and a
formal analogy may be very tempting. However, in the light of the above remarks,
it is clear  that such an analogy cannot be pushed too far: there is an essential difference between the two phenomena, namely, the presence of
strong correlations between velocity increments, leading to Kolmogorov's
 famous $k^{-5/3}$ law \ref{frisch}, which are absent in the case of
price increments \ref{comment}. As remarked above, the autocorrelation
function of price increments decreases rapidly to zero, resulting in 
a flat ("white-noise") spectrum (see figure below, taken from \ref{comment}). Translating into a power
spectrum we obtain a dependence well approximated by $\omega^{-2}$, which
merely reflects the absence of correlation, as opposed to the  $\omega^{-5/3}$
law for turbulent flows.

\psfig{file=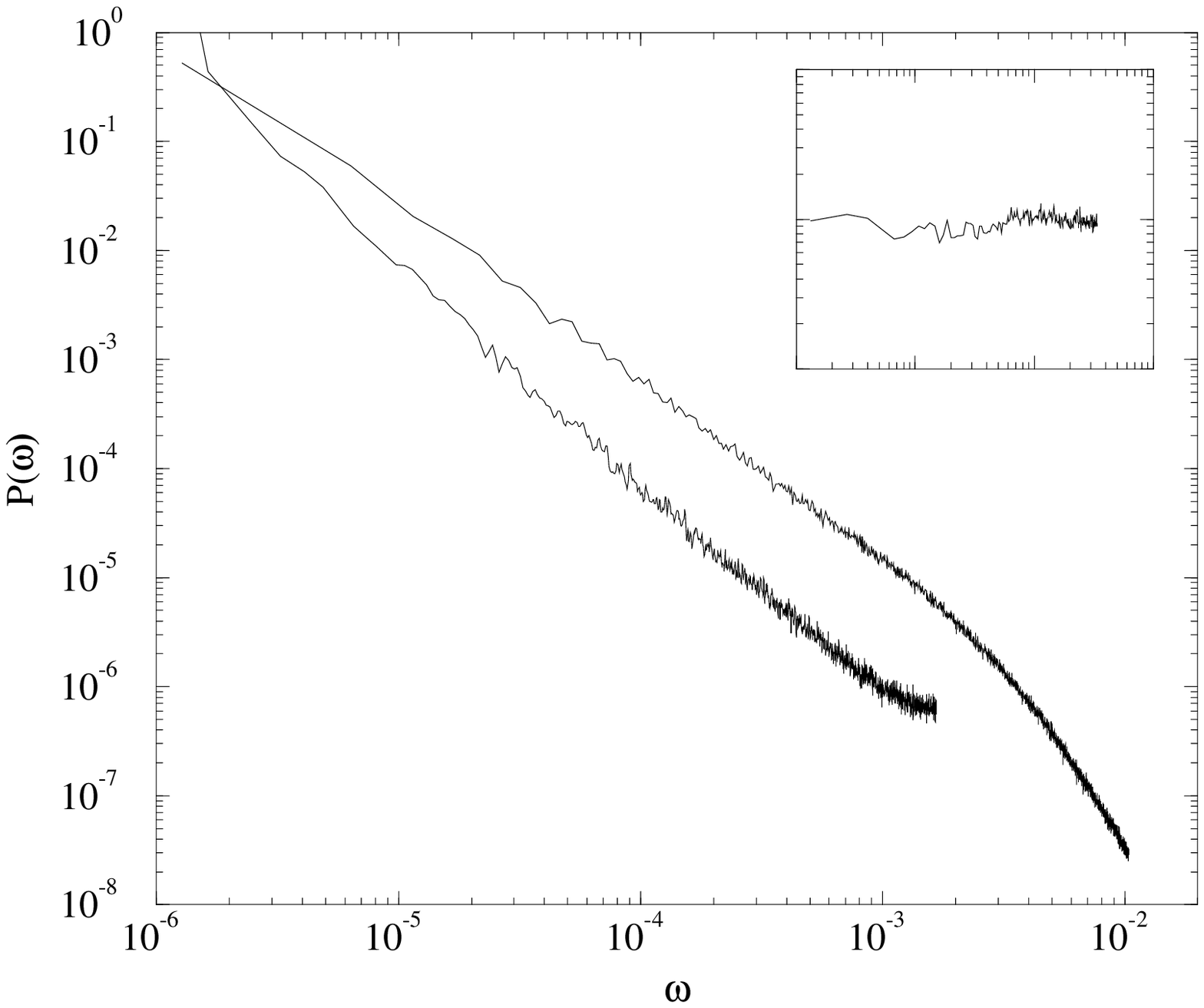,width=10cm,angle=0}
\caption{Fourier transform of the price autocorrelation function
$\langle x(t) x(t+\tau)\rangle$ as a function of temporal frequency for the
DEM-USD exchange rate (Oct 1991-Nov 1994) ({\it bottom curve}\/) as
compared to the $k^{-5/3}$ power-law spectrum observed for the
spatial velocity fluctuations in turbulent flows (the data were recorded by
Y. Gagne in a wind tunel experiment at $R_{\lambda}= 3050$) (top
curve). Inset: Fourier transform of the price change
autocorrelation function $\langle \Delta x(t) \Delta
x(t+\tau)\rangle$, which is completely flat (and would behave as the $1/3\ -th$ power of frequency in turbulent flows). Units are arbitrary.}

\section{Conclusion}

The above example illustrates that
although methods of statistical physics and in particular the concepts of scaling and scale invariance may be successfully applied in the study 
of market price fluctuations, sometimes revealing unsuspected  properties
of financial time series, analogies with problems in physics should be
handled with care in order to avoid erroneous conclusions. 
Keeping this point in mind, the scientific study of financial markets
proves to be a fascinating subject in itself, in particular for the physicist
whose theoretical tools may prove to be useful in uncovering new properties and mechanisms in financial data.

\vskip 1cm
\noindent
$\bullet$ Acknowledgements:

We thank J.P. Aguilar for his helpful remarks.
 
\references

\refis{comment}
 Arn\'eodo,A. {\it et al.} (1996)
"Comment on Turbulent Cascades in Foreign Exchange Markets" (preprint cond-mat/9607120) {\it Science \& Finance} Working Paper 96-01.

\refis{bachelier}
Bachelier, L. (1900) "Th\'eorie de la sp\'eculation "  
{\it Annales Scientifiques de l'Ecole Normale Sup\'erieure  }, 
{\bf III : 17}, 21-86. Bachelier's seminal work is also the first mathematical
study of Brownian motion, preceding Einstein's famous work on the same subject.

\refis{bouchaud}
Bouchaud, J.P. \& A. Georges (1990){\it Physics Reports}\ {\bf 195}, 125.

\refis{adapt}
Potters, M., Cont, R. \& Bouchaud, J.P. (1996) "Financial markets as adaptive ecosystems"
 (preprint cond-mat/9609172) {\it Science \& Finance} Working Paper 96-02.

\refis{book}
Bouchaud, J.P. \& Potters, M. (1997) {\it Th\'eorie des risques financiers} \ 
Paris: Al\'ea Saclay (forthcoming).

\refis{these}
Cont, R. (1997) {\it Dynamical models of financial markets} (Doctoral thesis), Universit\'e de Paris XI.

\refis{cont}
Cont, R. (1997) "Scaling and correlation in financial time series"
Science \& Finance Working Paper 97-01 (cond-mat/9705075).

\refis{fama1}
Fama, E.F. (1970) "Efficient capital markets: review of theory and empirical work"
{\it Journal of Finance},{\it 25}, 383-417.

\refis{fama2}
Fama, E.F. (1965) "The behavior of stock market prices"
{\it Journal of Business},{\it 38}, 34-105.

\refis{cootner}
Cootner, P. (ed.) {\it The random character of stock market prices}
Cambridge, MA: The MIT Press.

\refis{feller}
Feller, W. (1950) {\it Introduction to Probability theory and its applications},
{\bf II}, 3rd ed., John Wiley \& Sons, New York.

\refis{olsen2}
Guillaume, D.M. {\it et al.} "From the bird's eye to the microscope: a survey of new stylized facts of the intra-day foreign exchange markets" Olsen \& Associates Research Group working paper.

\refis{frisch} Frisch, U. (1995) {\it Turbulence: The Legacy of
A.N. Kolmogorov}, Cambridge University Press.

\refis{peinke} Ghashghaie, S. {\it et al.}(1996) "Tubulent cascades in foreign exchange markets" {\it Nature} {\bf 381} 767.

\refis{bs}
Black, F. \&  Scholes, M. (1973) "The pricing of options and corporate liabilities" {\it  Journal of Political Economy}, {\bf 81}, 635-654.

\refis{gnedenko}
Gnedenko, B.V. \& Kolmogorov, A.N. (1954)   
{\it Limit distributions for sums of independent random 
variables}, Addison-Wesley, Reading, MA.

\refis{koponen} Koponen, I. (1995) {\it Phys. Rev. E}\ {\bf 52}, 1197.

\refis{mandelbrot1}
Mandelbrot, B.  (1963) "The variation of certain speculative prices"
 {\it Journal of Business}, {\bf XXXVI}, 392-417. 

\refis{mandelbrot2}
Mandelbrot, B.  \& Taylor, H.M. (1967) "On the distribution of stock price differences " in {\it Operations Research},{\bf 15}, 1057-1062.

\refis{olsen}
Pictet O.V. {\it et al.} (1995) "Statistical study of foreign exchange rates, empirical evidence of a price change
scaling law and intraday analysis" {\it Journal of Banking and Finance}, {\bf 14}, 1189-1208.

\refis{pagan}
Pagan, A. (1996) "The econometrics of financial markets" {\it Journal of Empirical Finance},{\bf 3}, 15-102.

\refis{mantegna1}
Mantegna, R.N. (1991) "L\'evy walks and enhanced diffusion in the Milan stock exchange"
{\it Physica A}, {\bf 179}, 232.

\refis{mantegna2}
Mantegna, R.N. \& Stanley, H.E. (1995) "Scaling behavior of an economic index" {\it Nature}, {\bf 376}, 46-49. 

\refis{mantegna3}
Mantegna, R.N. \& Stanley, H.E.  (1994) {\it Phys. Rev. Lett.}, {\bf 73}, 2946.

\refis{zaj} 
Zajdenweber, D. (1994) "Propri\'et\'es autosimilaires du CAC40" 
{\it Revue d'Econo\-mie Politique}, {\bf 104}, 408-434.

\refis{santafe}
Pines, D. {\it et al.} (eds.) {\it The Economy as an evolving complex system}
Santa Fe Institute,  Addison-Wesley.

\refis{belkacem}
Belkacem, L. (1996) 
{\it Processus stables et applications \`a la finance}, Th\`ese de Doctorat,
 Universit\'e Paris IX.

\endreferences

\bye